# A Framework for Securing Health Information Using Blockchain in Cloud Hosted Cyber Physical Systems

Aisha Banu (1) , Sharon Priya S (1), Poojitha K(1), Kiruthiga R (1), R Annette  (2) Subash Chandran (3)

((1) BSA Crescent Institute of Science and Technology, (2) SM IEEE (3) NEC Corporation of America)

Electronic Health Records (EHRs) have undergone numerous technical improvements in recent years, including the incorporation of mobile devices with the cloud computing technologies to facilitate medical data exchanges between patients and the healthcare professionals. This cutting-edge architecture enables cyber physical systems housed in the cloud to provide healthcare services with minimal operational costs, high flexibility, security, and EHR accessibility. If patient health information is stored in the hospital database, there will always be a risk of intrusion, i.e., unauthorized file access and information modification by attackers. To address this concern, we propose a decentralized EHR system based on Blockchain technology. To facilitate secure EHR exchange across various patients and medical providers, we develop a reliable access control method based on smart contracts. We incorporate Cryptocurrency, specifically Ethereum, in the suggested system to protect sensitive health information from potential attackers.
In our suggested approach, both physicians and patients are required to be authenticated. Patients can register, and a block with a unique hash value will be generated. Once the patient discusses the disease with the physician, the physician can check the patient's condition and offer drugs. For experimental findings, we employ the public Block chain Ganache and solidity remix-based smart contracts to protect privacy. Ethers are used as the crypto currencies.

## 1. Introduction

Recently, there has been a rise in interest in the use of Blockchain technology to support medical and e-health services of cloud-hosted cyber physical systems. With its decentralized and trustworthy nature, Blockchain has demonstrated immense potential in a range of e-health disciplines, including the secure exchange of Electronic Health Records (EHRS) and data access management across numerous medical institutions [1]. Consequently, Blockchain adoption has the potential to provide innovative ways for increasing healthcare delivery and initiating a transformation of the healthcare business. Emerging technologies like Mobile Cloud Computing (MCC) and the Internet of Medical Things have facilitated major enhancements in E- health operations in the healthcare industry (IOMT). Patients can now obtain their medical records at home using mobile devices such as wearable sensors and or smart phones, share them with healthcare practitioners via cloud settings, which they can access quickly to analyze medical data and provide prompt medical care [2]. This revolutionary e-health service enables healthcare practitioners to track patients online and deliver ambulatory services at home, which not only improves the quality of care for patients but also reduces their expenses. In addition, having genuine EHRs in the cloud enables healthcare providers to monitor their patient's status and provide the proper diagnostic resources for diagnosis, treatment, and recovery. Despite these advantages, storing EHRs on servers frequently raises security problems, making cloud-based e-health systems challenging to deploy. The sharing of protected EHRs between patients and the providers of healthcare in mobile cloud environments is one of these security challenges. Without the approval of patients, unauthorized parties can gain harmful access to EHRs, jeopardizing the confidentiality, safety, and security of cloud-based e-health systems. As a result, it might be difficult for patients to monitor and manage their cloud-based health records, which are shared by healthcare providers. Consequently, mobile cloud EHR sharing systems require effective access management solutions. Traditional EHR access management solutions assume cloud servers are totally trusted by data owners, allowing them to implement both access control and authentication controls on data users. In mobile clouds, however, this paradigm is no longer applicable because the cloud server is dependable but untrustworthy. In the meanwhile, the cloud server will access confidential information without the users' authorization, which will result in serious data leakage and intrusion protection issues. In addition, standard access management systems emphasize a preset point of access, such as a centralized cloud service, which can result in data loss for e-health networks. Despite the fact that Blockchain-based access management or e-health offers a number of new authentication capabilities that are superior to those of conventional access control systems, these solutions are not yet widespread.

The rest of the chapter is organized as follows: Section 2 provides a detailed literature review, Section 3 gives an overview of the application of blockchain for security of medical data stored in cloud environment, Section 4 provides a detailed explanation about blockchain technology for the electronic health record system, its working methodology and the algorithm for applying blockchain to the electronic health record system. Section 5 explains in detail about the proposed blockchain architecture and finally the conclusion and future work are discussed in section 6.

## 2. Literature Survey

Blockchain technology is a decentralized knowledge exchange that is proprietary, free, trustworthy, and clear. The coordination and authentication activities are simplified in this situation because the documents are configured to refresh daily, and the two files are identical.

This article by Ahmad et al. [1] examines how blockchain technology solves scalability issues and offers alternatives in the healthcare sector. As a result, 16 technologies were grouped into two categories: storage management and blockchain renovation. The "Atlas. Ti" software was used to select the keywords and analyze the relevant articles. The six stages of this analysis include the following: Identification of the study topic, research methods, scanning of related articles, keywords based on the abstract, data extraction and the mapping process. The selected keywords were used to search through related papers using Atlas. ti tools. As a consequence, there are 48 codes and 403 quotes. As part of the mapping procedure, the codes are mapped onto the network as the next step. Notably, 16 methods fall into two categories: disc management and blockchain overhaul. In general, there are three storage optimization solutions and thirteen blockchain overhaul solutions, including read mechanism, blockchain modeling, write mechanism, and bi-directional network.

In the work of BL Radhakrishnan et al. [2], the patient's prescription information and medical history are stored in electronic health records. The attackers are drawn to the health reports because they contain useful material. The loss of an electronic health record may result in the administration of incorrect prescription or surgery. Healthcare programs have fewer compliance mechanisms in place to protect patient information.

Blockchain technology is a distributed and an open ledger that is critical for data and transaction security. The use of blockchain in healthcare networks defends patient information from hackers. Cold wallets, phishing, dictionary-based and hot wallets attacks are all threats to the blockchain. To secure the blockchain from the attacks described above, this paper proposes a multilevel authentication-based system. Machine Learning requires a reasonable amount of knowledge to form accurate decisions. In order to increase the accuracy of machine learning, data exchange and knowledge reliability are necessary. Blockchain technology's consensus ensures that data is authentic and safe. When those two technologies come together, machine learning can receive extremely precise leads thanks to the security and dependability of blockchain technology. The paper by Sonali Vyas et al. [3] summarizes how fusing these two technologies can benefit the medical and healthcare industries.

Rui Qiao et al. [4] Secure and trustworthy dynamic automated information that facilitates contact between the providers of healthcare will also increase the number of clinical research records, which is essential for expanding the breadth of medical trial cooperation, particularly in the case of uncommon diseases. Through a predetermined consent entry and consensus process, the consortium chain manages the confidentiality and traceability of clinical medical records. However, allowing people to own their medical data and securely and dynamically exchange it with other medical institutions remains a topic of interest. In addition, the intra-chain consortium consensus that supported the authentication node list is extended to the cross-chain consensus by modeling the cross-chain transaction consensus as a threshold digital signature mechanism with many privileged subgroups. According to testing results, the suggested method not only enables patients to communicate their information safely within milliseconds within a permitted medical consortium chain, but also enables complex interaction across heterogeneous consortium of chains.

Eman M Abou-Nassar et al. [5], Person, formal, and societal considerations currently place a premium on mobile and internet connectivity. Blockchains (BC), the Internet of Things (IoT), cloud networking, and other next-generation connection platforms offer limitless functionality for many applications and contexts, including as enterprises, cities, and the providers of healthcare. Sustainable convergence of the providers of healthcare nodes (devices, consumers, and suppliers, etc.) resulting in healthcare IoT (or IoHT) enables successful service delivery for the benefit of doctors, nurses, and so on. In federated confidence management systems, the semantic incompatibilities and a lack of suitable assets or attributes continue to impede secure knowledge exchange, whereas in IoHT, security, usability, and durability of medical data are of the utmost importance.

According to Abbas Yazdinejad et al. [6], interactions between the providers of healthcare to patients like the doctors, nurses, and other healthcare professionals must be secure and productive in any integrated healthcare system. In order to facilitate secure communications, numerous centralized authentication and security techniques outlined in the literature require the use of a respectable and trustworthy third party. Consequently, known overhead slows down authentication, patient flow, and interhospital communication. In this article, they suggest a blockchain-based technique for decentralized patient identification in a distributed hospital network. This enhancement would have a major influence on the network throughput, reduction of overhead, improvement in reaction time, and reduction in energy consumption. To demonstrate the broad viability of their proposed technique, they have also compared their model to a network's core concept without blockchain.

The exponential growth of cryptocurrencies and the blockchain technology underlying them have aided Szabo's original concept of smart contracts, which are electronic protocols that are designed to automatically facilitate, validate, and execute the negotiation and execution of all digital contracts without the need for central authorities. Smart contracts have been incorporated into popular blockchain-based development systems, such as Ethereum and Hyperledger, and have a vast array of deployment scenarios in the digital economy and intelligent markets, including management, healthcare, financial services and the Internet of Things, among others. As highlighted by Shuai Wang et al. [7], smart contracts are still in their infancy, and key technology issues such as protection and privacy remain unresolved.

According to Qalab E. Abbas et al. [8], the Role of blockchain in the intelligent mobile augmented reality project in particular is viewed as an encryption scheme to protect the privacy of the augmented objects and gives an overview of blockchain technology for this purpose.

Gagangeet Singh Aujla et al. [9] The health monitoring sensors constitute a huge IoT network that monitors the network continuously and transmits data to the nearby PCs and servers. The connectivity of these IoT-based sensors with other businesses, however, introduces security holes that an adversary can exploit due to the transparency of the data. This is a critical worry, especially in the healthcare industry, as changes in sensor data values might alter the direction of diagnosis, leading to severe health issues. In order to avoid data manipulation and ensure patient privacy, they provide a blockchain-based solution that is decoupled in the edge-envisioned future. This solution employs adjacent edge devices to establish decoupled blocks in the blockchain, enabling the secure transmission of healthcare data from sensors to edge nodes. This is intended to reduce data duplication in the expansive IoT healthcare network. The results demonstrate that the proposed approach is effective with regard to block preparation time, header generation time, tensor reduction ratio, and approximation error.

The present state of blockchain protocols for Internet of Things (IoT) networks is evaluated in this study [10] by Mohamed Amine Ferrag et al. They begin with a definition of blockchains and a summary of recent blockchain-related surveys. The authors then provide an outline of the implementation domains for blockchain technologies in IoT, such as the Internet of Energy, Internet of Vehicles, Internet of Cloud, Edge computing, etc. In addition, they categorize the threat types recognized by blockchain protocols in IoT networks into five categories: identity-based attacks, cryptanalytic attacks, manipulation-based attacks, service-based attacks and reputation-based attacks. In addition, they give a taxonomy and examination on a side-by-side of current options for stable and privacy-preserving blockchain solutions in terms of the blockchain paradigm, efficiency, basic security priorities, processing complexity, restrictions, and communication overhead. The results of the current study indicate that they address outstanding research questions and investigate potential future research topics in blockchain technology for IoT.

The other important facet of Electronic Health Records (EHRS) is that because of the nature of the system that involves collection of big data from various resources, it is inevitable to use the cloud resources for the storage of the information. The use of cloud hosted cyber systems has been extensively studied for various fields. Since the medical records are very sensitive any large the selection of the cloud services plays a vital role in ensuring safety and security of the resources which becomes a multicriteria problem and has been addressed by Ruby et al [11,12,13] in their works.

Sheng Cao et al. [14] The use of blockchain technologies (blockchain-based currencies, such as Ethereum) is suggested in this paper as a safe cloud-assisted e-Health infrastructure to shield outsourced EHRs from unlawful alteration. The core concept is that only authenticated parties can outsource EHRs and that each process on outsourcing EHRs is recorded as a transaction on the public blockchain. The modification of HER's cannot be completed until the subsequent transaction is documented into the blockchain and blockchain-based currencies offer a tamper proof way to execute transactions without central authority. As a result, any participant will verify the credibility of outsourced EHRs by looking at the subsequent transaction. The proposed system's security analysis and performance assessment show that it can have good security assurance while still being very effective.

In Healthcare a large amount of data is collected regularly to track patients, create medical reports, manage clinical trials and process medical insurance claims [15]. While the initial aim of blockchain implementations in use was to create distributed ledgers using virtual tokens, the momentum of this new technology has now spread to the medical field. With its adoption, it's critical to investigate how blockchain technology, combined with a smart contract structure, will help and threaten the healthcare domain across all interconnected players (patients, doctors, insurance providers, regulators) and involving properties (e.g., medical devices). Data from customers, physicians, the supply chain for supplies and drugs, and so on).

Leili Soltanisehat et al. [16] This article seeks to answer the following questions: What are blockchain's applications in healthcare systems, as well as the frameworks and challenges of how to integrate the blockchain to a specific healthcare domain. What is the temporal, technological, and spatial aspects of the blockchain implementations that are now being developed for various healthcare domains? What are the potential research avenues for the architecture and implementation of blockchain-based healthcare infrastructure? Statistical information regarding the technological dimensions of these 64 articles reveals that the majority of proposed blockchain-based healthcare networks utilize private Ethereum and blockchain platforms; additionally, the majority of authors are affiliated with academic institutions in the United States and China.

Akhilendra Pratap Singh et al. [17] In this study, a blockchain-based EHR and JavaScript-based Smart Contracts are integrated to create a decentralized, patient-centric healthcare management system. The proposed paradigm is secure due to the existence of a HyperLedger Fabric and Composer Technology-based working prototype. The results demonstrate the effectiveness of the proposed technique. Experiments conducted with the Hyper Ledger Calliper Benchmarking Tool illustrate the success of measures such as latency, throughput, resource utilization, etc. under a variety of conditions and management parameters.

Pronaya Bhattacharya et al. [18] This paper proposes a blockchain-based deep learning system as a service (BinDaaS). It employs deep learning and blockchain technologies to facilitate the two-phase exchange of EHR records amongst multiple healthcare users. In the first part, they describe an authentication and signature mechanism based on lattice-based encryption to avoid collusion attacks. In the second phase, the application of Deep Learning as a Service (DaaS) on stored data provides accurate disease diagnosis based on current patient characteristics and indicators.

## 3. Blockchain for security of medical data stored in cloud environment.

Cloud computing [19, 20, 21] has become one of the inevitable components of many health industry system architecture that are computationally very expensive. One of the new emerging areas in medicine is the use of animated models to explain the complex medical systems and concepts in medical education. The popularity of creating medical animations for medical education, research purposes and its usage has increased widely in recent years due to the availability of the high computational power at cheap prices using the cloud computing technology. These types of specialized cloud services required for medical animation files rendering are called the cloud render farm services [19, 20, 21] and these services provide the special high computational power and secure environments required for medical animation files rendering. Applying blockchain technology and encrypting the animation images from the user side makes it very difficult for security attacks and the medical animated models become more resistant to the security threats and attacks.

Cloud computing is also important for the data intensive architectures that scale up in course of time and the increase in the number of users like in the case of electronic health systems. Electronic health records are generated frequently for the same patient and the volume of the records generated over a small period of time is also very high. Thus, the cloud environment becomes inevitable to store these records. The safety of these records is also inevitable as these mostly contain various sensitive personal and health-related information. The selection of the right cloud service provider plays a vital role in ensuring the safety of the medical records stored in the cloud [22, 23]. However, using blockchain technology and encrypting the records from the user side makes it more resistant to security threats and attacks. The blockchain technology can also be used in the cloud computing domain to ensure the safe and secure storage of the files in the cloud environment [24, 25]. The bigger the data generated and requires processing, the higher is the computational speed required to process such files. Hence, the need of cloud services is inevitable in the medical domain industry to process high volume data like electronic health records and blockchain technology to make them secure in the cloud environment.

## 4. Blockchain for Electronic Health Record Systems.

Hospitals are increasingly using electronic health record (EHR) systems as a secure means of exchanging medical data. However, obtaining fragmented patient data through different EHRs continues to be a challenge because current EHRs are restricted to certain regions or are owned by affiliated institutions [26,27, 28]. The documentation of each incident that occurs when a medical record is viewed in an EHR system must be recorded in a log file for further auditing. The log file is a legal archive that may be utilized to recover the previous state of the patient's medical history. Therefore, they should take extreme precautions to prevent unauthorized access to the log file and, if at all feasible, make it unalterable.

The theft of medical data is still a possibility in the existing system because it is a centralized distribution system. The current framework is deficient in terms of network security, data privacy, and the dependability of sharing electronic health records across cloud servers. Furthermore, the existing system will always consist of a single point of failure, which will make the data unavailable. Disadvantages: There isn't enough security. Manual record administration, which is frequently subject to human influence and error. Both consumer trust and responsibility are lacking.

They proposed an EHR built on blockchain technology with a decentralized structure. To facilitate secure EHR exchange between patients and healthcare professionals, such as hospitals and pharmacists, they create a dependable access control system based on smart contracts. Patients can register for the proposed model and have access to their health records, which are then hashed into a single value and placed inside a QR code using the SHA 256 method. The doctor or hospital can access the patient's authorized records thanks to the hash value. A medication will be prescribed by the doctor, translated into a different block that the pharmacist can read, and an invoice will be created instantly. Advantages: Secure and tamper-proof. Smart contracts are used to preserve privacy. Agents acting as intermediaries are no longer required.

### 4.1 Blockchain Working Methodology:

Blockchain - A growing list of data entries that have been verified by all of the participating nodes are managed by the blockchain, a distributed database system. It employs methods now in use, including enlisting a trustworthy third party, to arrive at a system-wide consistent agreement without requiring cryptography, distributed computing, or game theory. A block is frequently a two-part data structure, with a header containing details about the hash value of the block before it, a timestamp, the Merkle root, and other metadata, and a data section containing transactional data. The hash values of preceding blocks are used to chain blocks together in chronological sequence. A group of nodes known as mining nodes maintains the whole chain of blocks and replicates it throughout the distributed network. Mining nodes also add new blocks to the chain using a predetermined consensus procedure. How unique a blockchain is compared to the rest of the network depends on its consensus algorithm. The Bitcoin protocol introduces proof-of-work, a consensus process dependent on each node's CPU power (PoW). Intelligent Contracts A "smart contract" is a set of goals and guiding principles for the creation of smart contracts. Digital technology can be used to provide safe communications through computer networks. Distributed ledgers are utilized in the execution of the contracts. Additionally, the execution processes are fully automated. This concept has been incorporated into a number of blockchain projects, most notably Ethereum and Hyperledger systems, as an extensive body of computer code that governs interactions between nodes in a network. The data is maintained on the blockchain network.

### 4.2 Algorithms for applying Blockchain:

Cryptography is the process of encrypting and decrypting data in order to use written code. A network that follows guidelines for block validation and nodal communication controls the blockchain. Transactions must be validated by miners before being recorded on the blockchain. Application of an algorithm is necessary for data validation and extraction in mining. Cryptocurrency is a type of digital money that employs cryptography to regulate and produce its unit pricing. Cryptocurrency is protected by encryption, and blockchain technology keeps track of transactions. The term "blockchain algorithm" refers to the complete procedure of adding records to the chain of transactions and verifying the Combinatorial algorithm. To be included in the blockchain, a block must satisfy a set of consensus requirements. Consensus algorithms are used in blockchain technology for this reason. The most widely used consensus method is known as Proof of Work (PoW). Because a blockchain network has numerous nodes or participants, the major goal of this method is to calculate any transaction that a participating node demands to be added to the network in order for it to operate. The nodes that do this activity are known as miners, and it is known as mining. Proof of Work (PoW), Proof of Stake (PoS), Proof of Burn (PoB), Practical Byzantine Fault Tolerance (PBFT), Proof of Capacity (PC), and Proof of Elapsed Time (PET) are a few examples of consensus algorithms. There are further strategies that foster consensus, including Leased Proof of Weight, Proof of Activity, Proof of Importance, Proof of Stake, and others. Blockchain networks cannot function without consensus algorithms that check every committed transaction, it is essential to choose a consensus algorithm carefully depending on the requirements of the business network. A cryptographic hash, like SHA-256, serves as the "signature" for a data file or text. SHA-256 creates a 256-bit (32-byte) signature for a text that is almost entirely unique. A hash is different from encryption because it cannot be reversed to disclose the original text because it is a one-way cryptographic function with a set size for any size of source text. As a result, rather than decrypting the text to retrieve the original version, it is acceptable to compare "hashed" versions of messages. Hash tables, integrity checks, challenge handshake authentication, digital signatures, and other tools are included in this collection of applications. When employing the "challenge handshake authentication" (also known as "challenge hash authentication") method, passwords are not sent in the "clear." Without having to worry that the server will discover the actual password, a client can transmit a password's hash over the internet for server validation. Communication becomes tamper-proof when the hash is linked to the original. The message can then be hashed once more by the receiver and compared to the hash that was provided; if they match, the message is unaltered. While creating a signature by encrypting a document's hash with your private key is more difficult with digital signatures, it is theoretically possible.

## 5. Modules in the Blockchain Architecture:

Internal Modules: Registration, Hash Value Generation, Block Generation, Privacy Preserving, Middleware, Smart Contract Creation. Patients and doctors who utilize the system can register by filling out the registration module. The registration module is where the user's data is gathered. The doctor must give the patient their name, phone number, email address, date of birth, address, and registration number. Each physician in the network is connected to a healthcare organization. The patient must give their name, contact information (phone, email, date of birth, and home address), as well as details about their health insurance. Furthermore, a work password is required from both patients and doctors. Making a smart contract is the process of writing a self-contained computer program that may be executed automatically. A smart contract on the Ethereum blockchain is nothing but a particular account that stores data and code with a variety of programmable features. Through application binary interfaces, users can interact with smart contracts using their Ethereum accounts. Smart contract functionalities can be activated by a fresh transaction received from an account. This property can be used by entities to carry out tasks including data transmission, request handling, and access management. A cryptographic hash is a technique for transforming an input into a fixed-size output. It looks to be made up of both letters and numbers. Cryptographic hashes can have many different shapes. Bitcoin uses the SHA-256 hashing method, for instance. An algorithm for hashing data is a computer function that shrinks the input data to a preset size and outputs a hash or hash value as the outcome. Data from files and words can be found, compared, and calculated using hashes. Before comparing the data to the original files, the program often computes a hash. When software is digitally signed and made available for download, this is a straightforward example of hashing. To achieve this, you'll need a hash of the script for the program you want to download. Additionally, a hashed digital signature is necessary. Once the input data has been hashed, the encrypted application can be downloaded. The browser must therefore decrypt the file and compare the two hash values when someone downloads software. The browser then performs a second hash on the file and the signature using the same hash function and method. The signature and the file can both be verified as authentic and unaltered if the browser creates the same hash value. Block creation – The patient must consent to the EHR communication. Double-spending attacks are avoided because of the patient's verification of the EHR. When the EHR is successfully verified, the Block Generator module constructs a block containing the EHR data. Each block of a blockchain stores the hash of the block data. The EHR is hashed using the key provided by the Key Generator Module. Maintaining privacy - Blockchain technology might, and ought to, be the answer for data security and privacy. The blocks are attached to one another in the form of a chain. The chain's first link is the Genesis block. Each block consists of a Block Header, Transaction Counter, and Transaction. The use of the patient's public address also safeguards the confidentiality of patient data. Instead of disclosing their private information, patients can share their public address with other members, such as doctors and hospitals. Middleware - The source of this dependency is a personal blockchain, which is a private development blockchain that may be used as a public blockchain. A tool for deploying and testing smart contracts is called Ganache. You have an operational personal blockchain network. Ten 100 Ether accounts are provided by Ganache so that our smart contracts can be tested on local blockchains. A type of digital asset that can be used as payment is cryptocurrency. Cryptocurrencies now have the security that is required to fend off attacks on the system's integrity and avoid double spending thanks to blockchain technology. When a user transfers cryptocurrency, there is no actual exchange of money. Instead, ownership is changed from the address of the sender to the address of the recipient. To access the coins being sent, the recipient needs the private key that decrypts the sender's public key. If the keys match, the transaction is added to the blockchain and simultaneously changes the balances in the sender's and receiver's addresses. The patient can pay bills, pharmacy costs, appointment prices, and medical expenditures among other things with cryptocurrency.

## 6. Conclusion

The blockchain technology is an important breakthrough but there are also some important drawbacks in the existing EHR sharing systems. The solutions to these issues have been discussed with the help of a working prototype. To provide effective and secure EHR sharing, this project aims to develop a dependable access control system based on a single smart contract to restrict user access. Our access control solution effectively detects and prevents unauthorized access to the e-health system, protecting network security and patient privacy. A significant quantity of storage is required for the blockchain-based healthcare system, which is still a difficult task. The future objective is to reduce the amount of storage needed for the storage of the blockchain is the objective of our future work. Since formal verification provides the highest level of assurance in smart contract behavior, it will become a crucial research area in the future.